\documentclass[final] {aipproc}
\usepackage{sidecap,graphicx,boxedminipage,epsfig,wrapfig,floatflt,shadow}

\layoutstyle{8x11double}

\begin{document}

\title{Rethinking Tools for Training Teaching Assistants}
\author{Chandralekha Singh}{
  address={Department of Physics and Astronomy, University of Pittsburgh, Pittsburgh, PA, 15260, USA}
}

\begin{abstract}
The ability to categorize problems is a measure of expertise in a domain. In order to help students learn effectively, 
instructors and teaching assistants (TAs) should have pedagogical content knowledge. They must be aware of the prior knowledge of students
they are teaching, consider the difficulty of the problems from students' perspective and design instruction that builds on what students 
already know. Here, we discuss the response of graduate students enrolled in a TA training course to categorization tasks 
in which they were asked to group problems based upon similarity of solution first from their own perspective, and later from the 
perspective of introductory physics students. Many graduate students performed an expert-like categorization of introductory physics problems. However, when asked to categorize the same problems from the perspective of introductory students, many graduate students expressed dismay, claiming that 
the task was impossible, pointless and had no relevance to their TA duties. 
We will discuss how categorization can be a useful tool for scaffolding and improving pedagogical content knowledge of
teaching assistants and instructors.
\end{abstract}

\maketitle

\section{Introduction}

The content knowledge of instructors is not sufficient to help students learn effectively. Indeed, instructors should possess pedagogical 
content knowledge and familiarize themselves with students' prior knowledge in order to scaffold their learning with appropriate
pedagogies and instructional tools.
Vygotsky's notion of ``zone of proximal development"~\cite{vygotsky} (ZPD) refers to what a student can do on his/her own vs. with the
help of an instructor who is familiar with his/her prior knowledge and skills.
Scaffolding is at the heart of ZPD and can be used to stretch a student's learning far beyond his/her initial knowledge by
carefully crafted instruction which is designed to ensure that the student makes desired progress and gradually develops independence.
With awareness of students' initial knowledge state, the instructor
can continuously target instruction a little bit above students' current knowledge state to ensure that the students
have the opportunity and ability to connect new knowledge with what they already know and build a robust knowledge structure. 

Piaget~\cite{piaget} emphasized ``optimal mismatch" between what the student knows and where the instruction should be targeted in order for 
desired assimilation and accommodation of knowledge to occur. 
Bransford and Schwartz~\cite{bransford} also proposed a framework for scaffolding student learning. They theorized that the preparation for 
future learning (PFL) and transfer of knowledge from the situation in which it was acquired to new situations is optimal if instruction includes both 
the elements of innovation and efficiency. In their model, efficiency and innovation are two orthogonal coordinates. If instruction only focuses
on efficiency, the cognitive engagement and processing by the students will be diminished and they will not develop the ability to transfer the
acquired knowledge to new situations. Similarly, if the
instruction is solely focused on innovation, students may struggle to connect what they are learning with their prior knowledge so that learning
and transfer will be inhibited. They propose that the preparation for future learning and transfer will be enhanced if the instruction focuses
on moving along a diagonal trajectory in the two dimensional space of innovation and efficiency.
One common element of all of these seemingly different frameworks is their focus on students' prior knowledge in order to scaffold learning.
Indeed, the instructor must be familiar with students' prior knowledge in order for 
instruction to be in the zone of proximal development and to provide optimal mismatch to ensure adequate preparation for future learning. 

A crucial difference between the problem solving strategies used by
experts in physics and beginning students lies in the interplay between
how their knowledge is organized and how it is retrieved to solve problems~\cite{fred,hardiman,intuition}.
In a classic study by Chi et al.\cite{chi3},
introductory physics students were asked to group mechanics problems into categories based
on the similarity of their solutions. 
Unlike graduate students (experts) who categorize them based on the
physical principles involved to solve them, introductory students
categorized problems involving inclined planes in one category and
pulleys in a separate category~\cite{chi3}.


Here, we will discuss the process and outcome of the categorization
of 25 introductory mechanics problems by 21 physics graduate students enrolled 
in a TA training course at the end of the course~\cite{epaps}. Graduate students
first performed the categorizations from their own perspective and
later from the perspective of a typical introductory student. 
We wanted to investigate if the graduate students have an understanding of the differences 
between their physics knowledge structure and those of the introductory physics students. 
One surprising finding is the resistance of graduate students to categorizing
problems from a typical introductory physics student's perspective
with the claim that such a task is ``useless", ``impossible", and has ``no
bearing" on their teaching assistant (TA) duties. Based on our finding, we suggest that
inclusion of such tasks can improve the effectiveness of TA training
courses and faculty development workshops and help TAs and instructors
focus on issues related to teaching and learning.

\vspace{-0.2in}
\section{Rating of Categories}
\vspace{-0.06in}

We were unable to obtain the questions in Ref.~\cite{chi3}
other than the few that have been published. We therefore chose
our own questions on sub-topics
similar to those chosen in Ref.~\cite{chi3}. The context of
the 25 mechanics problems varied and the topics included one- and
two-dimensional kinematics, dynamics, work-energy, and impulse-momentum~\cite{epaps}.
Many questions 
were adapted from an earlier study~\cite{chandra,singhvideo,edit} 
because their development 
had gone through rigorous testing.

Although we had an idea about which categories created by individuals should be considered good or
poor, we validated our assumptions with other experts. We randomly selected the categorizations performed 
by twenty introductory physics students and gave it to three physics faculty who had taught
introductory physics recently and asked them to decide whether each
of the categories created by individual students should be considered good,
moderate, or poor. We asked them to mark 
each row which had a category name created by a student and a description
of why it was the appropriate category for the questions that were
placed in that category. If a faculty member rated a category created by an introductory student as good,
we asked that he/she cross out the questions that did not belong to
that category. The agreement between
the ratings of different faculty members was better than 95\%.
We used their ratings as a guide to rate the categories created
by everybody as good, moderate, or poor. A category was considered ``good'' only if it was based
on the underlying physics principles. We typically
rated both conservation of energy or conservation
of mechanical energy as good categories. Kinetic
energy as a category name was considered a moderate
category if students did not explain that the questions placed in that
category can be solved using mechanical energy conservation or the
work energy theorem. We rated a category such as energy as good if students
explained the rationale for placing a problem in that category. If
a secondary category such as friction or tension was the only category
in which a problem was placed and the description of the
category did not explain the primary physics principles involved, it was considered
a moderate category.

More than one principle or concept
may be useful for solving a problem. The instruction for the categorizations told
students that they could place a problem in more than one category.
Because a given problem can be solved using more than one approach,
categorizations based on different methods of solution that are appropriate was considered good.
For some questions, conservation of mechanical energy may be more efficient,
but the questions can also be solved using one- or two-dimensional kinematics
for constant acceleration. In this paper, we will only discuss categories that
were rated good. If a graph shows
that 60\% of the questions were placed in a good category by a particular
group (introductory students, graduate students, or faculty), it means that the
other 40\% of the questions were placed in moderate or poor categories.


\vspace{-0.2in}
\section{Graduate students from Their Own Perspective}
\vspace{-0.06in}

A histogram of the percentage of questions placed in good
categories (not moderate or poor) is given in Fig.~1. This figure compares the average
performance of 21 graduate students at the end of a TA training course when they
were asked to categorize questions from their own perspective with
7 physics faculty and 180 introductory students who were given the
same task. Although this categorization by the graduate students is not on par with
the categorization by physics faculty, the graduate students displayed a higher
level of expertise in introductory mechanics than the introductory
students and were more likely to group the questions based on physical principles. 
\begin{figure}
\epsfig{file=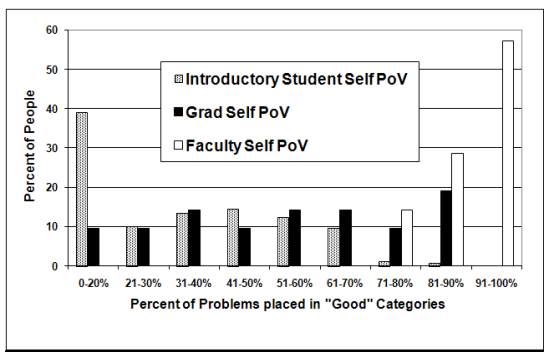,height=1.8in}
\caption{Histogram of percentages of introductory physics students, graduate students, and physics faculty who categorized various percentages
of the 25 problems in ``good" categories when asked to categorize them based upon similarity of solution
from their own point of view (Self PoV).}
\end{figure}
Physics professors and sometimes graduate students pointed out multiple methods for solving a problem
and specified multiple categories for a particular problem more often than the introductory students. Introductory students mostly placed one question
in only one category. Professors (and sometimes
graduate students) created secondary categories in which they placed a problem that were more like the introductory
students' primary categories. For example, in the questions
involving tension in a rope or frictional force~\cite{epaps}, many faculty and
some graduate students created these secondary categories called tension or friction, but also placed those questions in
a primary category, based on a fundamental principle of
physics. Introductory physics students were much more likely to place questions in inappropriate
categories than the faculty or graduate students, for example, placing a problem that was
based on the impulse-momentum theorem or conservation of momentum
in the conservation of energy category. 
Many of the categories generated by the three groups were the same, but
there was a major difference in the fraction of questions that were placed
in good categories by each group. 
There were some categories such as ramps, and pulleys, that were
made by introductory physics students but not by physics faculty or graduate students.

\vspace{-0.2in}
\section{Graduate students from Intro. Students' Perspective}
\vspace{-0.08in}

After the graduate students had submitted their own categorizations, they were asked
to categorize the same questions from
the perspective of a typical introductory physics student. A majority
of the graduate students had not only served as TAs for recitations, grading, or
laboratories, but had also worked during
their office hours with students one-on-one and in the Physics Resource Room at the University
of Pittsburgh. The goal of this task was to assess whether
the graduate students were familiar with the level of expertise of the introductory
students whom they were teaching and whether they realized that most introductory
students do not necessarily see the same underlying principles
in the questions that they do.
The graduate students were told that they were not expected to remember how
they used to think 4--5 years ago when they were introductory students. We wanted them to think about their experience as TAs in
introductory physics courses while grouping the questions from an introductory
students' perspective. They were also asked to specify whether they were recitation
TAs, graders, or laboratory TAs that semester.

The categorization of questions from the perspective of an introductory
physics student met with widespread resistance.
Many graduate students noted that the task was useless
or meaningless and had no relevance to their TA duties.
Although we did not tape record the discussion with the graduate students, we took
notes immediately following the discussion. The graduate students often asserted
that it is not their job to ``get into their students' heads.''
Other graduate students stated that the task was ``impossible'' and ``cannot
be accomplished.'' They often noted that they did not
see the utility of understanding the perspective of the students.
Some graduate students explicitly noted that the task was ``silly'' because
it required them to be able to read their students' minds and had
no bearing on their TA duties. Not a single graduate student stated that they saw
merit in the task or said anything in favor of why the task may
be relevant for a TA training course. The discussions with graduate students also suggest that many of them believed
that effective teaching merely involves knowing the content well and
delivering it lucidly. Many of them had never thought about the importance
of knowing what their students think for teaching to be effective.

It is surprising that most graduate students enrolled in the TA training course
were so reluctant or opposed to attempting the categorization task
from a typical introductory student's perspective. This resistance
is intriguing especially because the graduate students were given the task at the
end of a TA training course and most of them were TAs for introductory
physics all term. It is true that it is very difficult for the TAs
(and instructors in general) to imagine themselves as novices.
However, it is possible for TAs
(and instructors) to familiarize themselves with students'
level of expertise by giving them pre-tests at the beginning of a
course, listening to them carefully, and by reading literature
about student difficulties, for example, as part of the TA training course.

After 15--20 minutes of discussion we made the task more concrete
and told graduate students that they could consider categorizing from the perspective
of a relative whom they knew well after he/she took only one introductory
mechanics course if that was the only exposure to the material they
had. We also told them that they had to make a good faith effort even
if they felt the task was meaningless or impossible. 
Figure~2 shows the histogram of how the graduate students categorized questions from their own perspective
and from the perspective of a typical introductory student/relative
who has taken only one physics course and also categorization by introductory students. 
Figure~2 shows that the graduate students 
re-categorized the questions in worse categories when performing the categorization from the perspective
of a typical introductory physics student.
However, if we look at questions placed in each category, for example, conservation
of momentum, there are sometimes significant differences between the
categorization by graduate students from an introductory students' perspective
and by introductory students from their own perspective. This implies that while graduate students may have
realized that a typical introductory student/relative who has taken only one physics course may not perform as well as a physics graduate
student on the categorization task, overall they were not able to anticipate the frequency with which introductory students
categorized each problem in the common less-expert-like categories.
\begin{figure}[h!]
\epsfig{file=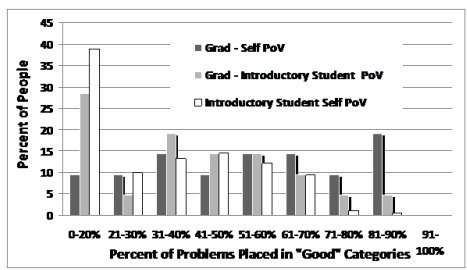,height=1.4in} 
\caption{Histogram of percentages of introductory students and graduate students who categorized various
percentages of the 25 problems in ``good'' categories when
asked to categorize them based on similarity of solution. Graduate students categorized from their
own point of view and from the perspective of a typical introductory physics student. 
}
\end{figure}

\vspace{-0.2in}
\section{Discussion and Summary}
\vspace{-0.09in}

The reluctance of TAs to re-categorize the questions from introductory students' perspective 
raises the question of what should the graduate students learn in a TA training class. In a typical TA training class, a significant
amount of time is devoted to emphasizing the importance of writing clearly on the blackboard, speaking
clearly and looking into students' eyes, and grading students' work
fairly. There is a lack of discussion about the fact that teaching
requires not only knowing the content but understanding how students
think and implementing strategies that are commensurate with students'
prior knowledge.

After the graduate students had completed both sets of categorization tasks, we
discussed the pedagogical aspects of perceiving and evaluating
the difficulty of the questions from the introductory students' perspective.
We discussed that pedagogical content knowledge, which is critical
for effective teaching, depends not only on the content knowledge
of the instructor, but also on the knowledge of what the students
are thinking. The discussions were useful and many students
explicitly noted that they had not pondered why accounting for
the level of expertise and thinking of their students was important for devising strategies to facilitate learning. Some graduate students noted that they will listen to their introductory
students and read their responses carefully.

One graduate student noted that after this discussion he felt that, similar to the
difficulty of the introductory students in categorizing the introductory
physics questions, he has difficulty in categorizing questions in the
advanced courses he has been taking. He added that when he is assigned
homework/exam questions, for example, in the graduate level electricity and
magnetism course in which they were using the classic book by Jackson,
he often does not know how the questions relate to the material discussed
in the class even when he carefully goes through his class notes. The student noted that
if he goes to his graduate course instructor for hints, the instructor
seems to have no difficulty making those connections to the homework.
The spontaneity of the instructor's connection to the lecture material and the
insights into those questions suggested to the student that the instructor
can categorize those graduate-level questions and explain the method
for solving them without much effort. This facility is due in part because
the instructor has already worked out the questions and hence they have become an exercise. Other graduate
students agreed with his comments saying they too had similar experiences
and found it difficult to figure out how the concepts learned in the
graduate courses were applicable to homework problems assigned in
the courses. These comments are consistent with the fact that a graduate
student may be an expert in the introductory physics material related
to electricity and magnetism but not necessarily an expert in the
material at the Jackson level course. 

This study raises important issues regarding the content of TA training
courses and faculty professional development workshops and the extent
to which these courses should allocate time to help participants learn
about pedagogical content knowledge in addition to the usual discussions
of logistical issues related to teaching. Asking the graduate students and faculty to categorize questions from the
perspective of students may be one way to draw instructor's attention to these important issues
in the TA training courses and faculty professional development workshops.

\vspace*{-.23in}
\begin{theacknowledgments}
We thank J. Brascher for help in data analysis and NSF for 
awards NSF-PHY-0653129 and PHY-055434.

\end{theacknowledgments}
\vspace*{-.11in}

\bibliographystyle{aipproc}

\end{document}